\documentclass[aps,prd,twocolumn,showpacs,amsmath,amssymb]{revtex4-1}
\usepackage{amsmath} \usepackage{graphicx} \usepackage{subfigure}
\usepackage{epstopdf} \usepackage{color} \usepackage{multirow}
\usepackage{setspace} \usepackage{overpic} \usepackage{amssymb}
\usepackage{ulem}
\usepackage{siunitx}

\usepackage[bookmarksnumbered, pdfstartview=FitH,colorlinks,urlcolor=blue, citecolor=blue,linkcolor=blue] {hyperref}
\usepackage{lineno}
\usepackage{bm}
\usepackage{rotating}
\usepackage{xcolor}
\usepackage{makecell}
\usepackage{mathrsfs}
\usepackage[utf8]{inputenc}
\usepackage{threeparttable}

\let\oldequation\equation
\let\oldendequation\endequation
\renewenvironment{equation}
 {\linenomathNonumbers\oldequation}
 {\oldendequation\endlinenomath}

\include{def-com-new}
\begin{document}

\title{\boldmath Search for $\chi_{c1} \to \pi^+\pi^-\eta_c$ via $\psi(3686)\to \gamma \chi_{c1}$ }

\author{
\begin{small}
\begin{center}
M.~Ablikim$^{1}$, M.~N.~Achasov$^{4,c}$, P.~Adlarson$^{77}$, X.~C.~Ai$^{82}$, R.~Aliberti$^{36}$, A.~Amoroso$^{76A,76C}$, Q.~An$^{73,59,a}$, Y.~Bai$^{58}$, O.~Bakina$^{37}$, Y.~Ban$^{47,h}$
% ...
\\
\vspace{0.2cm}
(BESIII Collaboration)\\
\vspace{0.2cm} {\it
$^{1}$ Institute of High Energy Physics, Beijing 100049, People's Republic of China\\
% ...
$^{4}$ Budker Institute of Nuclear Physics SB RAS (BINP), Novosibirsk 630090, Russia\\
% ...
$^{36}$ Johannes Gutenberg University of Mainz, Johann-Joachim-Becher-Weg 45, D-55099 Mainz, Germany\\
$^{37}$ Joint Institute for Nuclear Research, 141980 Dubna, Moscow region, Russia\\
% ...
$^{47}$ Peking University, Beijing 100871, People's Republic of China\\
% ...
$^{58}$ Southeast University, Nanjing 211100, People's Republic of China\\
$^{59}$ State Key Laboratory of Particle Detection and Electronics, Beijing 100049, Hefei 230026, People's Republic of China\\
% ...
$^{73}$ University of Science and Technology of China, Hefei 230026, People's Republic of China\\
% ...
$^{76}$ University of Turin and INFN, (A)University of Turin, I-10125, Turin, Italy; (B)University of Eastern Piedmont, I-15121, Alessandria, Italy; (C)INFN, I-10125, Turin, Italy\\
$^{77}$ Uppsala University, Box 516, SE-75120 Uppsala, Sweden\\
% ...
$^{82}$ Zhengzhou University, Zhengzhou 450001, People's Republic of China\\
\vspace{0.2cm}
$^{a}$ Deceased\\
% ...
$^{c}$ Also at the Novosibirsk State University, Novosibirsk, 630090, Russia\\
% ...
$^{h}$ Also at State Key Laboratory of Nuclear Physics and Technology, Peking University, Beijing 100871, People's Republic of China\\
% ...
}\end{center}
\vspace{0.4cm}
\end{small}
}

\date{\today}

\def\BBF{\ensuremath{3.1 \times10^{-4}}}
\def\Totsys{\ensuremath{6.6}}

\begin{abstract}
Utilizing $(2712.4 \pm 14.3) \times 10^6$ $\psi(3686)$ events collected with the BESIII detector at the BEPCII collider, we search for the hadronic transition process $\chi_{c1} \to \pi^+\pi^-\eta_c$ following the decay $\psi(3686)\to \gamma \chi_{c1}$. No significant signal is observed, and an  upper limit of $\mathcal{B}(\chi_{c1}\to\pi^+\pi^-\eta_c)$ is determined to be \BBF~at 90\% confidence level, which is one order of magnitude more stringent than the previous measurement. 
\end{abstract}

\maketitle
\section{Introduction}

Quarkonium states, consisting of a quark bound to an anti-quark ($q\bar{q}$), have provided good laboratories for understanding the fundamental theory of the strong interactions between quarks and gluons~\cite{quarkonia}. 
The hadronic transitions involving heavy quarkonium states emitting a pion pair, and isospin-violating transitions emitting a single neutral pion, have garnered a significant and lasting interest due to their relevance in understanding the quantum chromodynamics~(QCD) of heavy and light hadrons.  In particular, the QCD multipole expansion~(QCDME)~\cite{qcdme1,qcdme2,qcdme3} has made successful predictions for many hadronic transition rates in the $c\bar{c}$ and $b\bar{b}$ systems ~\cite{qcdme4,qcdme5,PRD054019,PRD074033}.

Among charmonium transitions from the $\chi_{cJ}$ states to $\eta_c$ with emission of one or two pions, only the decays $\chi_{c1} \to \pi^+\pi^-\eta_c$ and $\chi_{c0} \to \pi^0\eta_c$ arise from the leading E1-M1 term of the QCDME and are thus most likely to be observed~\cite{PRD074033}.
Voloshin~\cite{PRD074033} also predicts that $\mathcal{B}(\chi_{c0} \to \pi^0\eta_c) \approx \mathcal{B}(\chi_{c1} \to \pi^+\pi^-\eta_c)$.  The BESIII collaboration has set upper limits on these branching fractions (BFs) of $\mathcal{B}(\chi_{c1} \to \pi^+\pi^-\eta_c) < 3.2\times 10^{-3}$~\cite{PRD012002} and $\mathcal{B}(\chi_{c0} \to \pi^0\eta_c) < 1.6\times 10^{-3}$~\cite{PRD112018}, at 90\% confidence level~(CL), utilizing a dataset of $106 \times 10^6~ \psi(3686)$ events~\cite{CPC023001}. 
The upper limit of $\mathcal{B}(\chi_{c1} \to \pi^+\pi^-\eta_c)$ is approximately one order of magnitude below a prediction based on E1-M1 transitions: $\mathcal{B}(\chi_{c1} \to \pi^+\pi^-\eta_c)$ = (2.22 $\pm$ 1.24)\%~\cite{PRD054019}. This indicates that further study on these decays is needed. 

Recently, the upper limit $\mathcal{B}(h_c \to \pi^0J/\psi) <4.7 \times 10^{-4}$ at 90\% CL has been obtained with data samples collected at center-of-mass energies between 4.189 and 4.437 GeV by BESIII~\cite{JHEP003}. The $\mathcal{B}(\chi_{c1} \to \pi^+\pi^-\eta_c)$ is expected to be less than $1.1 \times 10^{-4}$ under the assumption of $\Gamma(\chi_{c0} \to \pi^0\eta_c) \approx  3\Gamma(h_c \to \pi^0J/\psi)$ and $\mathcal{B}(\chi_{c0} \to \pi^0\eta_c) \approx \mathcal{B}(\chi_{c1} \to \pi^+\pi^-\eta_c)$~\cite{PRD074033}. 


Based on a data sample of $(2712.4 \pm 14.3) \times 10^6 ~\psi(3686)$ events collected by the BESIII experiment~\cite{CPC023001, psi2S2021}, approximately $264.4 \times 10^6$ $\chi_{c1}$ particles are expected to be produced via the radiative transition process $\psi(3686)\to\gamma \chi_{c1}$. With such a substantial signal yield, there is a great chance to investigate the hadronic transition $\chi_{c1} \to \pi^+\pi^-\eta_c$ by fully reconstructing the \(\eta_c\) through 16 dominant decay channels~\cite{pdg2024}. 

\section{BESIII DETECTOR AND MONTE CARLO SIMULATION}
\label{sec:BES}

The BESIII detector~\cite{Ablikim:2009aa} records symmetric $e^+ e^-$ collisions provided by the BEPCII storage ring~\cite{CXYu_bes3} in the center-of-mass energy range from 1.84 to 4.95~GeV,
with a peak luminosity of $1.1 \times 10^{33}\;\text{cm}^{-2}\text{s}^{-1}$
achieved at $\sqrt{s} = 3.773\;\text{GeV}$. BESIII has collected large data samples in this energy region~\cite{Ablikim:2019hff,EcmsMea,EventFilter}. The cylindrical core of the BESIII detector covers 93\% of the full solid angle and consists of a helium-based multilayer drift chamber (MDC), a plastic scintillator time-of-flight system (TOF), and a CsI(Tl) electromagnetic calorimeter (EMC), which are all enclosed in a superconducting solenoidal magnet providing a 1.0~T magnetic field. The solenoid is supported by an octagonal flux-return yoke with modules of resistive plate muon counters interleaved with steel. The charged-particle momentum resolution at 1~GeV/c is 0.5\%, and the  d$E/$d$x$ resolution is 6\% for the electrons from Bhabha scattering. The EMC measures photon energy with a resolution of 2.5\% (5\%) at 1~GeV in the barrel (end-cap) region. The time resolution in the plastic scintillator TOF barrel region is 68~ps, while that of the end-cap part is 110~ps. The end-cap TOF system was upgraded in 2015 using multi-gap resistive plate chamber technology, providing a time resolution of 60~ps, which benefits about $83\%$ of the data used in this paper~\cite{tof_a,tof_b,tof_c}.

Monte Carlo (MC) simulated data samples produced with a {\sc geant4}-based~\cite{geant4} software package, which includes the geometric description of the BESIII detector and the detector response, are used to optimize the event selection criteria, and to estimate the signal efficiency and background level. The simulation models the beam-energy spread and initial-state radiation in the $e^+e^-$ annihilation using the generator {\sc kkmc}~\cite{kkmc_a,kkmc_b}. The inclusive MC sample includes the production of the $\psi(3686)$ resonance, the initial-state radiation production of the $J/\psi$, and the continuum processes incorporated in {\sc kkmc}~\cite{kkmc_a,kkmc_b}. Particle decays are generated by {\sc evtgen}~\cite{evtgen_a,evtgen_b} for the known decay modes with BFs taken from the Particle Data Group~(PDG)~\cite{pdg2022} and by {\sc lundcharm}~\cite{lundcharm_a,lundcharm_b} for the unknown ones. Final-state radiation from charged final-state particles is included using the {\sc photos} package~\cite{photos}.

A total of 270,000 signal MC events are generated for each of the 16 exclusive hadronic decay channels of $\eta_c$, divided up between the three different data-taking periods, and corresponding detector simulation, based on the total number of $\psi(3686)$ events in each. Here, the signal MC events of the $\psi(3686) \to \gamma \chi_{c1}$ decays follow the angular distribution of $1 -\frac{1}{3} \cos^2\theta_{\gamma}$, where $\theta_{\gamma}$ denotes the polar angle of the radiative photon~\cite{evtgen_a,evtgen_b}, while the other sub-decays are simulated by a phase space model.

\section{EVENT SELECTION}
\label{sec:selection}

The $\eta_c$ is reconstructed from the 16 decay channels detailed in Table~\ref{tab:eff}. Therefore, the final-state particles in this analysis consist of $\pi, K, p, \pi^0, \eta, K_{S}^0$, and $\gamma$. Charged tracks detected in the MDC are required to be within a polar angle ($\theta$) range of $\vert\!\cos\theta\vert<0.93$, where $\theta$ is defined with respect to the $z$ axis, which is the symmetry axis of the MDC.  For each charged track, time-of-flight and d$E$/d$x $ information are combined to compute a particle identification (PID) $\chi^2_{\rm PID}$ for the pion, kaon, and proton hypotheses.  This PID information is used for best-candidate selection only, as described later.  
For charged tracks not originating from $K_S^0$ decays, the distance of closest approach to the interaction point (IP) must be less than 10\,cm along the $z$ axis, $|V_{z}|$, and less than 1\,cm in the transverse plane.

\begin{table}[htbp]\footnotesize
	\caption{The BFs, $\mathcal{B}_i$, and selection efficiencies, $\epsilon^{\rm sig}_i$, for the 16 $\eta_c$ decay channels, $X_i$).  
The summed branching fraction times efficiency, accounting $\pi^0, \eta, K^0_S$ BFs not included here, is 2.36\%. }
	\centering
	\begin{tabular}{ccc}
	\hline
	\hline
		$X_i$ &$\mathcal{B}_i$~(\%)~\cite{PRD092009}  &$\epsilon^{\rm sig}_i$ (\%)\\
	\hline
		$p\bar{p}$ 					&0.15 $\pm$ 0.04    &29.60\\
		$K^+K^-\pi^+\pi^-$ 				&0.95 $\pm$ 0.23   &13.72\\
		$p\bar{p}\pi^+\pi^-$ 				&0.53 $\pm$ 0.19   &12.20\\
		$\pi^+\pi^-\pi^+\pi^-$     			&1.72 $\pm$ 0.36   &16.84\\
		$K^+K^-K^+K^-$ 				&0.22 $\pm$ 0.09   &11.86\\		
		$\pi^+\pi^-\pi^+\pi^-\pi^+\pi^-$    	&2.02 $\pm$ 0.54   &6.73\\
		$K^+K^-\pi^+\pi^-\pi^+\pi^-$    		&0.83 $\pm$ 0.24   &4.32\\
		$K_S^0K^{\pm}\pi^{\mp}$ 		&2.60 $\pm$ 0.51   &14.57\\		
		$K_S^0K^{\pm}\pi^{\mp}\pi^{\pm}\pi^{\mp}$    		&2.75 $\pm$   0.74 &5.73\\
		$K^+K^-\pi^0$ 					&1.04 $\pm$  0.23 &13.23\\
		$p\bar{p}\pi^0$ 				&0.35 $\pm$0.14     &13.69\\
		$K^+K^-\eta$ 					&0.48 $\pm$  0.25   &12.21\\
		$\pi^+\pi^-\eta$     				&1.66 $\pm$  0.46 &14.59\\
		$\pi^+\pi^-\pi^0\pi^0$     			&4.66 $\pm$ 1.01   &7.89\\
		$\pi^+\pi^-\pi^+\pi^-\eta$     		&4.40 $\pm$ 1.28  &6.71\\
		$\pi^+\pi^-\pi^+\pi^-\pi^0\pi^0$     	&17.23 $\pm$  3.30 &2.87\\
        \hline
	\hline
	\end{tabular}
	\label{tab:eff}
\end{table}

Good photon candidates are identified using showers in the EMC.  The deposited energy of each shower must be more than 25~MeV in the barrel region ($\vert\!\cos\theta\vert< 0.80$) and more than 50~MeV in the end-cap region ($0.86 <\vert\!\cos\theta\vert< 0.92$).  
To suppress electronic noise and showers unrelated to the event, the difference between the EMC time and the event start time is required to be less than 700\,ns.

Neutral pion and $\eta$ candidates are reconstructed by combining two good photons, then a one-constraint (1C) kinematic fit is applied, with the invariant mass of $\gamma \gamma$ constrained to the known $\pi^0$ or $\eta$ mass~\cite{pdg2024}, and a requirement of $\chi^2_{1 \rm C} < 200$ is applied.

Each $K_{S}^0$ candidate is reconstructed from two oppositely charged tracks satisfying $|V_{z}|<$ 20~cm. The two charged tracks are assigned as $\pi^+\pi^-$ without imposing further particle identification~(PID) criteria. They are constrained to originate from a common vertex and are required to have an invariant mass within $|M_{\pi^{+}\pi^{-}} - m_{K_{S}^{0}}|<$ 12~MeV$/c^{2}$, where $m_{K_{S}^{0}}$ is the known $K^0_{S}$ mass~\cite{pdg2024}. The decay length of the $K^0_S$ candidate is required to be greater than twice the vertex resolution away from the IP.  The updated $K_S^0$ four-vector is used for later kinematics.  

The $\eta_c$ candidates are reconstructed in the 16 exclusive decay modes, $X_i$, and the events are accepted or rejected based on their consistency with the $\chi_{c1} \to \pi^+\pi^- \eta_c, \eta_c \to X_i$ hypothesis. Events containing at least one combination meeting the criteria of $p_{\pi^\pm} < 0.4$ GeV/$c$ for the momentum of the two soft pions, and with the recoil mass of $\gamma$ and the two soft pions, RM$(\gamma \pi^+\pi^-)$, within the range [2.80, $3.20$] GeV/$c^2$ are retained for further analysis.

To suppress background, a four-constraint (4C) kinematic fit is performed, constraining the four-momentum of the final state particles to that of the initial system. The $M_{\gamma\gamma}$ mass resolution is improved by the 4C kinematic fit. We require the $\gamma\gamma$ combinations to satisfy  $0.101 < M_{\gamma \gamma} < 0.153$ GeV/$c^2$ for $\pi^0$ candidates and $0.524 < M_{\gamma \gamma} < 0.566$ GeV/$c^2$ for $\eta$ candidates. 
In cases where multiple $\eta_c$ candidates are identified in an event, the one with the smallest value of $\chi^2 = \chi^2_{4 \rm C} + \chi^2_{1 \rm C} + \chi^2_{\rm PID} + \chi^2_{\rm vertex}$ is selected. Here, $\chi^2_{1 \rm C}$ represents the (summed) $\chi^2$ value(s) from the 1C kinematic fit(s) of the $\pi^0(\eta)$, $\chi^2_{\rm PID}$ is the sum of PID $\chi^2$ values for all charged tracks, and $\chi^2_{\rm vertex}$ is the $\chi^2$ from the $K_S^0$ vertex fit. If events do not involve $\pi^0, \eta$, or $K_S^0$, the corresponding $\chi^2_{1 \rm C}$ or $\chi^2_{\rm vertex}$ is set to zero.

Considering the resolution difference between different $\eta_c$ decay channels, these decay channels are classified into charged~(C), $K_S^0$~(K) and neutral~(N) categories. 
The C channel includes $p\bar{p}$, $K^+K^-\pi^+\pi^-$, $p\bar{p}\pi^+\pi^-$, $\pi^+\pi^-\pi^+\pi^-$, $\pi^+\pi^-\pi^+\pi^-\pi^+\pi^-$, $K^+K^-\pi^+\pi^-\pi^+\pi^-$ and  $K^+K^-K^+K^-$.
The K channel consists of $K_S^0K^{\pm}\pi^{\mp}$ and $K_S^0K^{\pm}\pi^{\mp}\pi^{\pm}\pi^{\mp}$. 
The N channel includes $K^+K^-\pi^0$, $p\bar{p}\pi^0$, $K^+K^-\eta$, $\pi^+\pi^-\eta, \pi^+\pi^-\pi^0\pi^0$, $\pi^+\pi^-\pi^+\pi^-\eta$ and  $\pi^+\pi^-\pi^+\pi^-\pi^0\pi^0$.
To suppress background, we optimize selection criteria for $\chi^2_{4 \rm C}$ by maximizing a figure-of-merit~\cite{ref:fom} given by $\epsilon_{\rm sig}/(\alpha/2+\sqrt{B})$. Here the signal efficiency, $\epsilon_{\rm sig}$, is determined with the signal MC sample, the background yield, $B$, is estimated by the inclusive MC sample, and $\alpha$ represents the significance level, which is set to be 3.0. 
The optimal requirements of $\chi^2_{4\rm C}$ are found to be less than 42, 36, and 23 for the C, K, and N decay channels, respectively. 


To suppress background from the decays $\psi(3686) \to \pi^+\pi^- J/\psi, J/\psi \to \gamma \eta_c$ and $\psi(3686) \to \eta J/\psi, \eta \to \gamma \pi^+\pi^-, J/\psi \to \gamma \eta_c$, we require that the recoil mass of the $\pi^+\pi^-$, RM($\pi^+\pi^-$), satisfy $|$RM($\pi^+\pi^-) - m_{J/\psi}| > 3 \sigma_{J/\psi}$, and the invariant mass of $\gamma \pi^+\pi^-$, $M(\gamma \pi^+\pi^-)$, satisfy $|M(\gamma \pi^+\pi^-) - m_{\eta}| > 3 \sigma_{\eta}$. 
Here, $m_{J/\psi}$ and $m_{\eta}$ represent the known masses of the $J/\psi$ and $m_{\eta}$~\cite{pdg2024}, while their respective mass resolutions are $\sigma_{J/\psi} = 2.43$ MeV and $\sigma_{\eta} = 4.62$ MeV, obtained from fits to the signal MC sample.

\begin{figure}[htbp]
\begin{center}
\begin{minipage}[t]{1\linewidth}
\subfigure[\label{sacfig:a}]{\includegraphics[width=0.49\textwidth]{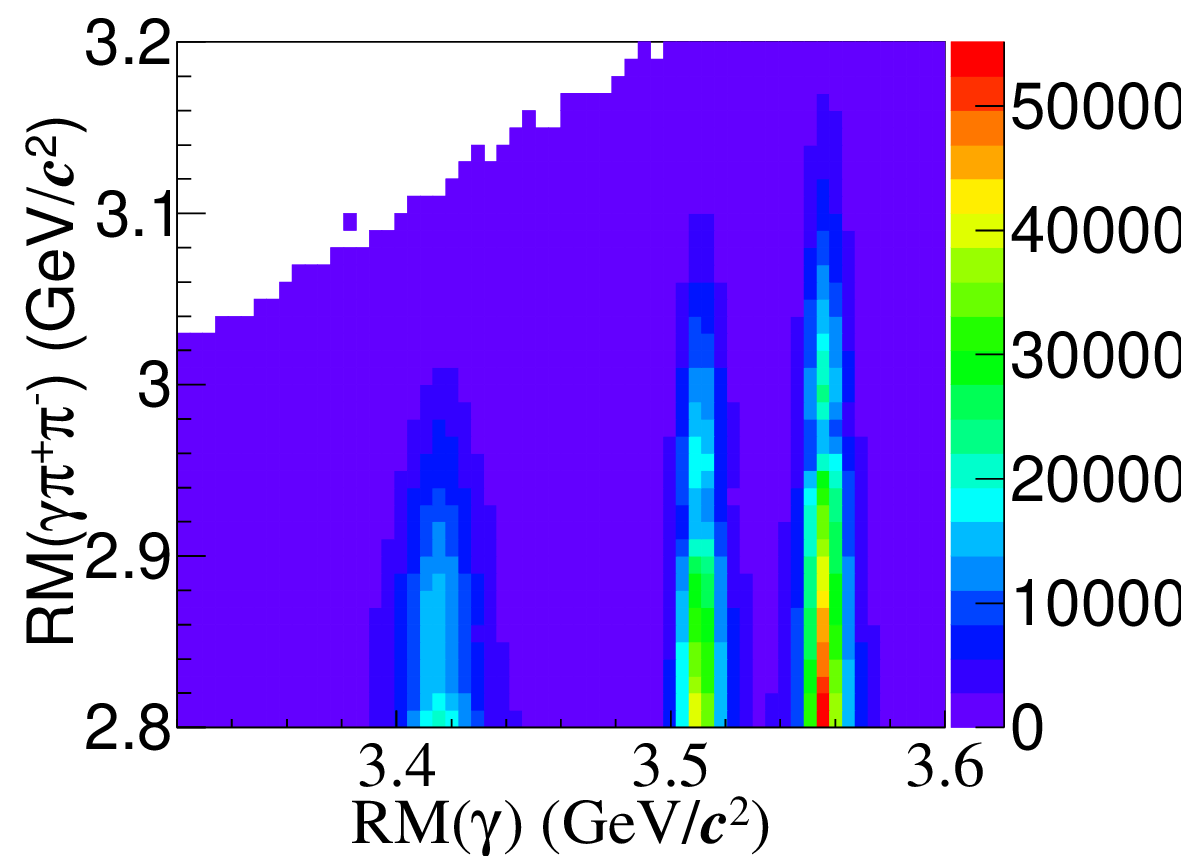}}
\subfigure[\label{sacfig:b}]{\includegraphics[width=0.49\textwidth]{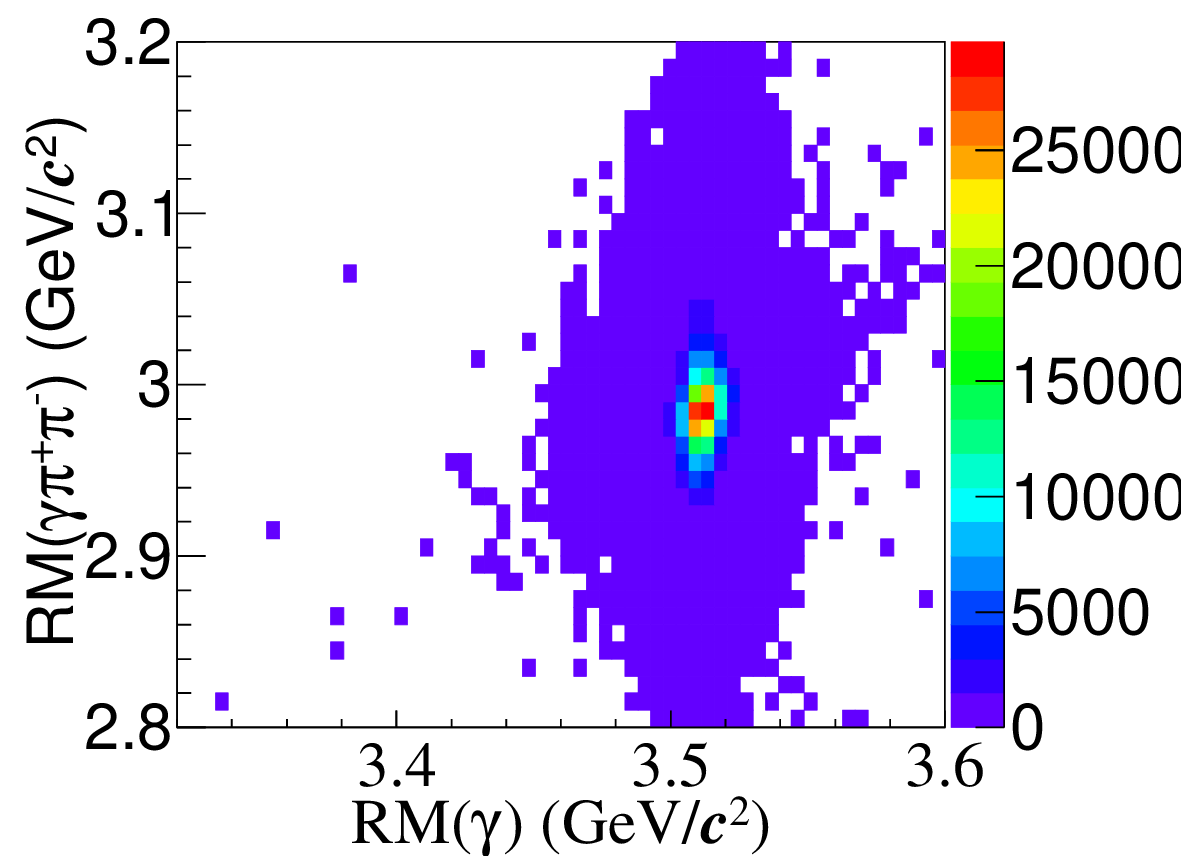}}
\end{minipage}
\caption{Distributions of the recoil masses, RM($\gamma \pi^+\pi^-$) vs. RM($\gamma$), for (a) data sample and (b) signal MC sample.} 
\label{fig:fit}
\end{center}
\end{figure}

Following the background analysis and selection criteria discussed above, the remaining background events share the same final state as the signal process and exhibit a smooth RM($\gamma \pi^+\pi^-$) distribution. The distributions of RM($\gamma \pi^+\pi^-$) versus RM($\gamma$) for data and signal MC sample are shown in Fig.~\ref{sacfig:a} and Fig.~\ref{sacfig:b}, respectively.

\section{\label{Sec:BR_determined}Signal yields}
We select the $\chi_{c1}$ signal events within the mass region [3.48, 3.54] GeV/$c^2$, which corresponds to about 5$\sigma_{\chi_{c1}}$ based on a fit to the signal MC sample. Subsequently, we perform a maximum likelihood fit to the RM($\gamma\pi^+\pi^-$) distribution to extract $\eta_c$ signal events. In the absence of clear $\eta_c$ signal events, we combine the 16 $\eta_c$ decay modes and fit to all accepted candidates from the data, as shown in Fig.~\ref{fig:fit}. 
The combinatorial background shape is segmented into two parts for characterization: the dip as shown in Fig.~\ref{fig:fit} attributed to a RM($\pi^+\pi^-$) cut is described by a Johnson function~\cite{Johnson}, while the smooth component is modeled by a fifth-order polynomial. 
The $\eta_c$ signal is described using the shape obtained from signal MC sample. 
The signal yield determined from the fit is $N^{\rm{obs}}_{\rm{sig}} = -2819 \pm 1653$. We also perform a fit on the accepted candidates for the inclusive MC samples, excluding signal events, and verify that the fit procedure does not introduce any bias.
The BF is calculated as
\begin{align}
\label{equp}
&\mathcal{B}(\chi_{c1} \to \pi^+\pi^-\eta_c) = \nonumber \\ 
&\frac{N_{\rm sig}}{N_{\psi(3686)}\; \mathcal{B}(\psi(3686) \to \gamma \chi_{c1}) \sum_{i} \left(\mathcal{B}_i \; \mathcal{B}_{\rm sub}(X_i) \; \epsilon^{\rm sig}_i\right)}.
\end{align}
Here, $N_{\rm sig}$ denotes the signal yield and $N_{\psi(3686)}$ is the total number of $\psi(3686)$ events. The BF and detection efficiency of the $i$-th signal mode are denoted as $\mathcal{B}_{i}$ and $\epsilon^{\rm sig}_i$, as listed in Table~\ref{tab:eff}. The quantities $\mathcal{B}_{\rm sub}(X_i)$ represent the BFs of the intermediate states, including $\mathcal{B}(\pi^0 \to \gamma \gamma)$, $\mathcal{B}(\eta \to \gamma \gamma)$, and $\mathcal{B}(K_S^0 \to \pi^+\pi^-)$, quoted from the PDG~\cite{pdg2024}. 
Due to the absence of a signal in the fit, 
an upper limit $N_{\rm sig}^{\rm up} = 1713$  at 90\% CL is obtained from the Bayesian method with a step size of $N_{\rm sig} = 1.0$ and a prior density of $p=0$ for $N_{\rm sig}<0$, and $p=1$ for $N_{\rm sig}\geq0$. 
After incorporating the systematic uncertainties addressed in Sec.~\ref{sec:sysU}, we obtain $N_{\rm sig}^{\rm up} = 2016$ and the upper limit on $\mathcal{B}(\chi_{c1} \to \pi^+\pi^-\eta_c)$ is calculated to be \BBF  ~at 90\% CL.

 \begin{figure}[htbp]
\begin{center}
\begin{minipage}[t]{1\linewidth}
\includegraphics[width=1\textwidth]{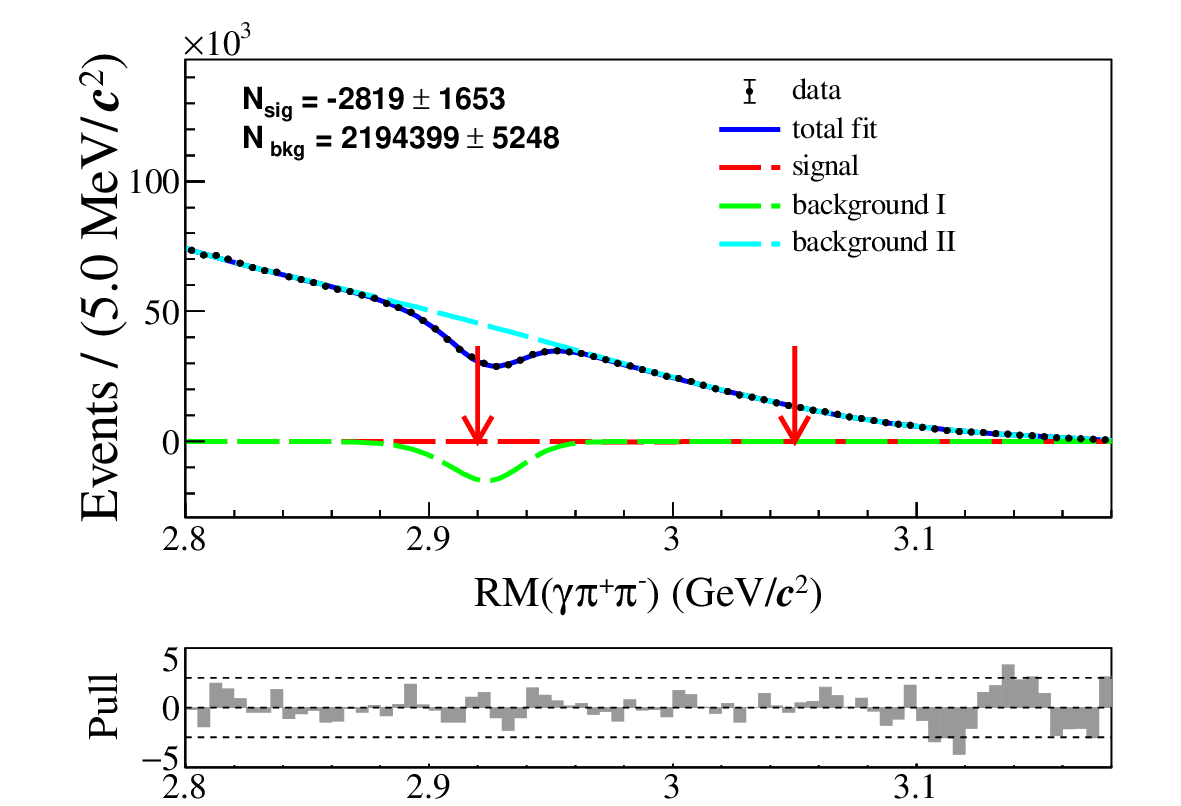}
\end{minipage}
\caption{The fit to the recoil mass RM$(\gamma\pi^+\pi^-)$ distribution. The points with error bars are data, the solid blue curve is the fit result, the dashed red curve is the signal, and the dashed cyan curve is a combinatorial background shape represented by a fifth-order polynomial.  The dashed green curve is a Johnson function~\cite{Johnson} which models a dip induced by an RM($\pi^+\pi^-$) requirement.  
The red arrows are intended solely to indicate the anticipated primary distribution of the signal events, no specific criteria have been established for the signal window.} 
\label{fig:fit}
\end{center}
\end{figure}

\section{Systematic Uncertainties}\label{sec:sysU}
The systematic uncertainty for the upper limit of  the BF includes multiplicative and additive sources.  The multiplicative systematic uncertainties are from the total number of $\psi(3686)$ events, tracking, photon detection, kinematic fitting, quoted BFs, mass windows, and the model used to simulate the $\eta_c$ decays.  Details of the systematic uncertainties are discussed below.

\subsection{Uncertainty related to the input values}
\begin{itemize}
\item[(i)]{{\bf Total number of $\psi(3686)$ events}}: 
The total number of $\psi(3686)$ events in data is determined by using inclusive hadronic $\psi(3686)$ events with an uncertainty of 0.5\%~\cite{CPC023001,psi2S2021}.

\item[(ii)]{\bf {Quoted BFs}}:

The $\mathcal{B}(\psi(3686)\to \gamma \chi_{c1}) = (9.75\pm0.27)\%$ is taken from the PDG~\cite{pdg2024} and the corresponding uncertainty, 2.8\%, is used. The $\mathcal{B}(\pi^0 \to \gamma \gamma) = (98.82\pm0.03)\%$, $\mathcal{B}(\eta \to \gamma \gamma) = (39.36\pm0.18)\%$ and $\mathcal{B}(K_S^0 \to \pi^+ \pi^-) = (69.20\pm0.05)\%$ are also quoted from the PDG~\cite{pdg2024} and the corresponding uncertainties are neglected. 
The $\mathcal{B}(\eta_c\to X_i)$ are taken from the previous BESIII measurement~\cite{PRD092009}, as listed in Table~\ref{tab:eff}.
The corresponding uncertainty on $\mathcal{B}(\eta_c\to X_i)$ is calculated to be 7.2\% by
 \begin{eqnarray}\begin{aligned}
 \label{eq:sys_un}
\frac{\sqrt{\sum_{i=1}^{16} \left( \epsilon^{\rm sig}_i \; \mathcal{B}_i \; \delta_i \right)^2}}{\sum_{i=1}^{16} \epsilon^{\rm sig}_i \; \mathcal{B}_i}, 
\end{aligned}\end{eqnarray}
\end{itemize}
where $\delta_i$ is the fractional uncertainty on $\mathcal{B}_i$. 

\subsection{Uncertainty related to the efficiency}

\begin{itemize}
\item[(i)]{{\bf Tracking}}:
The uncertainty due to charged particle tracking is determined to be 1.0\% per track by using the control samples of $J/\psi \to \pi^+\pi^-\pi ^0, J/\psi \to p\bar{p}\pi^+\pi^-, J/\psi \to K_S^0K^+\pi^- + c.c.$~\cite{PRD112005}.

\item[(ii)]{\bf Photon detection}: 
Using the control samples $J/\psi \to \rho^0\pi^0$ and $e^+e^- \to \gamma \gamma$~\cite{PRD052005}, the uncertainly due to photon reconstruction is assigned as 1.0\% per photon.

\item[(iii)]{\bf PID efficiency}: 
The PID uncertainty is determined to be 1.0\% per track~\cite{PRD112005}, by using the control samples of $J/\psi \to \pi^+\pi^-\pi ^0, J/\psi \to p\bar{p}\pi^+\pi^-, J/\psi \to K_S^0K^+\pi^- + c.c$. 

\item[(iv)]{\bf $\bm{K_S^0}$ reconstruction}: 
Based on the studies using the control sample $J/\psi \to K^{*\pm}K^{\mp}$, the uncertainly due to $K_S^0$ reconstruction is assigned as 1.2\% per $K_S^0$.

\item[(v)]{\bf $\bm{\pi^{0}, \eta}$ reconstruction}:
Using a high-purity control sample of $J/\psi \to \pi^0p\bar{p}$, the systematic uncertainty from the reconstruction is determined to be 1.0\% per $\pi^0$ or $\eta$~\cite{PRL261801}.

\item[(vi)]{\bf Kinematic fit}:
The efficiencies $\epsilon^{\rm sig}_i$ are re-obtained from the signal MC sample after correcting the helix parameters of the charged tracks.  Half of the difference in the efficiencies with and without this correction~\cite{YPG:bam} is taken as the associated uncertainty.

\item[(vii)]{\bf Mass windows}:
A Barlow test~\cite{barlow_test} is performed to examine the deviation in significance, $\zeta$, between the baseline selection and that performed for the systematic test. The deviation in significance is defined as
\begin{equation}
    \zeta=\frac{\left|V_{\text {nominal }}-V_{\text {test }}\right|}{\sqrt{\left|\sigma_{V \text { nominal }}^2-\sigma_{V \text { test }}^2\right|}},
\end{equation}
where $V$ represents $N_{\rm sig}^{\rm up}/\epsilon^{\rm sig}_i$ and $\sigma_{V}$ is the statistical uncertainty on $V$. The boundaries of the mass windows applied on RM$(\pi^+\pi^-)$, $M(\gamma \pi^+\pi^-)$ and RM$(\gamma)$ are varied by $\pm 1 \sigma$, with $\sigma$ obtained from the fit to the MC samples. As the $\zeta$ values are always less than 2, the corresponding systematic uncertainties are ignored. 

\item[(viii)]{\bf $\bm{\eta_c}$ MC models}:
We employ a phase space model to simulate the decays of $\eta_c$ To quantify the systematic uncertainties arising from the omission of intermediate states in these decays, a study of the control sample of $\psi(3686)\to\gamma \eta_c, \eta_c \to X_i$ has been conducted~\cite{PRD092009}. The differences in efficiencies between the phase space and alternative MC simulations accounting for observed substructure in the decays are taken as the systematic uncertainties.

\item[(ix)]{\bf MC statistics}:
The binomial uncertainty on efficiencies due to MC statistics is calculated.  
\end{itemize}

\begin {table*}[htbp]\small
	\caption{Relative systematic uncertainties, in \%, related to efficiency of the $\eta_c$ exclusive decay channels. 
Uncertainties from tracking, photon detection, PID, $K_S^0, \pi^0$, $\eta$ reconstruction and kinematic fit are summed using Eq.~(\ref{eq:sys_co}), reflecting to their correlations, while those from the $\eta_c$ decay models and MC statistics are summed using Eq.~(\ref{eq:sys_un}).}
	\centering
	\begin{tabular}{c|c|c|c|c|c|c|c|c}
	\hline
	\hline
		$X_i$  &Tracking	&Photon	&PID	&$K_S^0$	&$\pi^0~(\eta)$	&Kinematic fit	&$\eta_c$ MC model	&MC statistics\\
	\hline
		$p\bar{p}$ 					    &4.0 &-	&4.0	&-	&-	&2.4	&0.0			&0.5\\
		$K^+K^-\pi^+\pi^-$ 				   &6.0 &-	&6.0	&-	&-	&2.9	&0.6			&0.8\\
		$p\bar{p}\pi^+\pi^-$ 				   &6.0 &-	&6.0	&-	&-	&4.2 &2.5			&0.8\\
		$\pi^+\pi^-\pi^+\pi^-$     			   &6.0 &-	&6.0	&-	&-	&0.6	&2.1			&0.7\\
		$K^+K^-K^+K^-$     				 &6.0 &-	&6.0	&-	&-	&3.0		&0.5		&0.8\\
		$\pi^+\pi^-\pi^+\pi^-\pi^+\pi^-$    	   &8.0 &-	&8.0	&-	&-	&5.7		&0.0		&1.1\\
		$K^+K^-\pi^+\pi^-\pi^+\pi^-$    		   &8.0 &-	&8.0	&-	&-	&2.2		&3.0		&1.4\\
		$K_S^0K^{\pm}\pi^{\mp}$     		 &4.0 &-	&4.0	&1.2	&-	&1.7		&0.5		&0.8\\
		$K_S^0K^{\pm}\pi^{\mp}\pi^{\pm}\pi^{\mp}$     &6.0 &-	&6.0	&1.2	&-	&0.8	&5.2		&1.3\\
		$K^+K^-\pi^0$ 					  &4.0 &2.0	&4.0	&-	&1.0	&5.0		&4.6		&0.8\\
		$p\bar{p}\pi^0$ 				     &4.0 &2.0	&4.0	&-	&1.0	&7.4		&5.8		&0.8\\
		$K^+K^-\eta$ 					   &4.0 &2.0	&4.0	&-	&1.0	&4.5		&8.1		&0.8\\
		$\pi^+\pi^-\eta$     				 &4.0 &2.0	&4.0	&-	&1.0	&2.4		&5.5		&0.8\\
		$\pi^+\pi^-\pi^0\pi^0$     			  &4.0 &4.0	&4.0	&-	&2.0	&1.7		&0.1		&1.1\\
		$\pi^+\pi^-\pi^+\pi^-\eta$     		  &6.0 &2.0	&6.0	&-	&1.0	&2.5		&0.0		&1.1\\
		$\pi^+\pi^-\pi^+\pi^-\pi^0\pi^0$     	 &6.0 &4.0	&6.0	&-	&2.0	&5.2		&0.5		&1.8\\
        \hline
        \hline
       Sum	&5.3 &1.8	&5.3	&0.2	&0.9	&3.0		&0.5			&0.4\\
        \hline
	\hline
	\end{tabular}
	\label{tab:totalsys}
\end{table*}

The systematic uncertainties from the tracking, photon detection, PID, $K_S^0, \pi^0$, $\eta$ reconstruction and kinematic fit are correlated between each decay mode; the sum of the 16 decay channels is thus calculated by 
 \begin{eqnarray}\begin{aligned}
\frac{\sum_{i=1}^{16} \left(\epsilon^{\rm sig}_i \; \mathcal{B}_i \; \delta_i\right)}{\sum_{i=1}^{16} \epsilon^{\rm sig}_i \;\mathcal{B}_i}, 
\label{eq:sys_co}
\end{aligned}\end{eqnarray}
where $\delta_i$ is the mode-dependent uncertainty.  
The systematic uncertainties from the $\eta_c$ decay models and MC statistics are uncorrelated, and the sum of the 16 decay channels is calculated with Eq.~(\ref{eq:sys_un}).

The combined systematic uncertainty on the efficiency of $\eta_c\to X_i$ is 8.3\%, summing the uncertainties from all relevant sources in quadrature. All multiplicative systematic uncertainties are summarized in Table~\ref{tab:all-Sys}; the total, 11.3\%, is obtained by adding them in quadrature assuming that they are independent.

The additive systematic uncertainties stem from the determination of the signal yield, which depends on the fit range and the shapes used to describe the signal and background contributions. Since these additive systematic uncertainties are correlated, we simultaneously change these three fit conditions and choose the most conservative upper limit. The variations used to evaluate these systematic uncertainties are discussed below.

\begin{itemize}
\item[\textbullet]  {\bf Fit range}:
The systematic uncertainty due to the fit range is examined by enlarging or shrinking the fit range by $\pm$5 MeV/$c^2$, $\pm$10 MeV/$c^2$, $\pm$15 MeV/$c^2$, and $\pm$20 MeV/$c^2$, given the mass resolution on RM($\gamma\pi^+\pi^-$) 
of about 5 MeV/$c^2$. 

\item[\textbullet]  {\bf Signal shape}:
The systematic uncertainty from the signal shape is estimated by using the MC-simulated shape convolved with a Gaussian function, $G(\delta m, \delta \sigma)$. Here, $\delta m = (-2.31\pm0.05$) MeV and $\delta \sigma = (3.65\pm0.24$) MeV are taken from the control sample of \(\psi(3686) \to \pi^+ \pi^- J/\psi\), \(J/\psi \to \gamma \eta_c\). 

  \item[\textbullet]  {\bf Background shape}:
The systematic uncertainty due to the background shape is estimated by replacing the fifth-order polynomial and a Johnson function with a sixth-order polynomial and a Novosibirsk function. 
\end{itemize}

\begin{figure}[htbp]
\begin{center}
\begin{minipage}[t]{0.9\linewidth}
\includegraphics[width=1\textwidth]{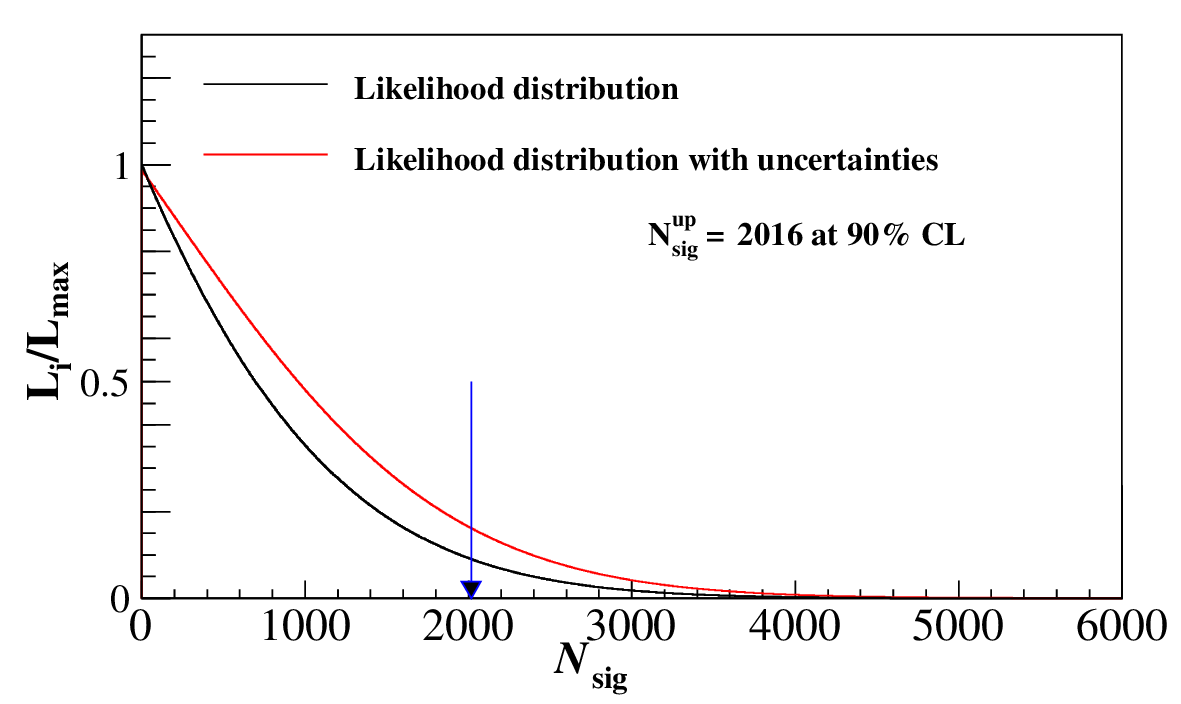}
\end{minipage}
\caption{The normalized likelihood distribution for the signal yield of $\psi(3686) \to \gamma \chi_{c1}, \chi_{c1} \to \pi^+\pi^-\eta_c$. The curve obtained without any systematic uncertainties is depicted in black, 
and another considering all systematic uncertainties is in red.  The arrow indicates the upper limit on the signal yield at 90\% CL from the latter curve.} 
\label{fig:newup}
\end{center}
\end{figure}

After adopting the most conservative upper limit by accounting for the additive systematic uncertainties, the multiplicative systematic is incorporated into the calculation of the upper limit via~\cite{Stenson:2006gwf, Liu:2015uha}
\begin{equation}
   L'(N_{\rm sig}) = \int_0^1 {L({\textstyle{\epsilon \over {\epsilon_{\rm sig}}}}N_{\rm sig})} \exp \left[ { - {\textstyle{{(\epsilon - \epsilon_{\rm sig})^2} \over {2\sigma _{\epsilon}^2}}}} \right]d\epsilon,
\end{equation}
where $L(N_{\rm sig})$ is the the original likelihood distribution as a function of signal yield, $N_{\rm sig}$, $\epsilon$  is the expected efficiency and $\epsilon_{\rm sig}$ represents the combined efficiency based on the BFs of the 16 decay channels, $\sigma_{\epsilon}$ is the multiplicative systematic uncertainty as summarized in Table~\ref{tab:all-Sys}. The obtained likelihood distribution is shown in Fig.~\ref{fig:newup}. 
Taking into account the systematic uncertainties, the upper-limit signal yield for $\psi(3686) \to \gamma \chi_{c1},  \chi_{c1} \to \pi^+\pi^-\eta_c$ at 90\% CL is $N^{\rm up}_{\rm sig}$ = 2016 is used for determining the upper limit BF.

\begin{table}[htbp]
  \caption{Summary of multiplicative systematic uncertainties.}
  \label{tab:all-Sys}
  \begin{center}
    \begin{tabular}{cccc}
      \hline
      \hline
      Source   &Uncertainty (\%)\\
      \hline
      Number of $\psi(3686)$ events               & 0.5 \\
      $\mathcal{B}(\psi(3686)\to \gamma \chi_{c1})$     & 2.8\\            
      $\mathcal{B}(\eta_c\to X_i)$     & 7.2 \\
      $\mathcal{B}(\pi^0 \to \gamma \gamma)$	&Negligible\\
      $\mathcal{B}(\eta \to \gamma \gamma)$		&Negligible\\
      $\mathcal{B}(K_S^0 \to \pi^+ \pi^-)$		&Negligible\\
      Efficiency          &8.3\\
      \hline
      Total                               &11.3 \\
      \hline
      \hline      
    \end{tabular}
  \end{center}
\end{table}

\section{Summary}
A search for the hadronic transition $\chi_{c1} \to \pi^+\pi^-\eta_c$ via $\psi(3686)\to \gamma \chi_{c1}$ has been performed. No significant signal is observed, and the upper limit on $\mathcal{B}(\chi_{c1} \to \pi^+\pi^-\eta_c)$ at 90\% CL is determined to be \BBF. Assuming isospin symmetry and neglecting the small phase space difference between charged and neutral $\pi \pi$ modes yields a combined limit of $\mathcal{B}(\chi_{c1} \to \pi\pi\eta_c) < 4.6\times 10^{-4}$, at 90\% CL.  

The result obtained in this paper is one order of magnitude lower than 90\% CL upper limit of $\mathcal{B}(\chi_{c1} \to \pi^+\pi^-\eta_c) < 3.2 \times 10^{-3}$ set by the  previous BESIII study~\cite{PRD012002}.
The experimental measurements are compatible with the theoretical prediction of Voloshin~\cite{PRD074033} that $\mathcal{B}(\chi_{c0} \to \pi^0\eta_c) \approx \mathcal{B}(\chi_{c1} \to \pi^+\pi^-\eta_c)$. 


\textbf{Acknowledgement}

The BESIII Collaboration thanks the staff of BEPCII (https://cstr.cn/31109.02.BEPC) and the IHEP computing center for their strong support. This work is supported in part by National Key R\&D Program of China under Contracts Nos. 2020YFA0406300, 2020YFA0406400, 2023YFA1606000, 2023YFA1606704; National Natural Science Foundation of China (NSFC) under Contracts Nos. 12375070, 11635010, 11935015, 11935016, 11935018, 12025502, 12035009, 12035013, 12061131003, 12192260, 12192261, 12192262, 12192263, 12192264, 12192265, 12221005, 12225509, 12235017, 12361141819; the Chinese Academy of Sciences (CAS) Large-Scale Scientific Facility Program; Joint Large Scale Scientific Facility Funds of the NSFC and CAS under Contracts Nos. U2032108; CAS under Contract No. YSBR-101; Shanghai Leading Talent Program of Eastern Talent Plan under Contract No. JLH5913002; 100 Talents Program of CAS;  The Institute of Nuclear and Particle Physics (INPAC) and Shanghai Key Laboratory for Particle Physics and Cosmology; German Research Foundation DFG under Contract No. FOR5327; Istituto Nazionale di Fisica Nucleare, Italy; Knut and Alice Wallenberg Foundation under Contracts Nos. 2021.0174, 2021.0299; Ministry of Development of Turkey under Contract No. DPT2006K-120470; National Research Foundation of Korea under Contract No. NRF-2022R1A2C1092335; National Science and Technology fund of Mongolia; National Science Research and Innovation Fund (NSRF) via the Program Management Unit for Human Resources \& Institutional Development, Research and Innovation of Thailand under Contract No. B50G670107; Polish National Science Centre under Contract No. 2024/53/B/ST2/00975; Swedish Research Council under Contract No. 2019.04595; U. S. Department of Energy under Contract No. DE-FG02-05ER41374.



\begin{thebibliography}{99}


  
  \bibitem{quarkonia} 
N. Brambilla \textit{et al.} [Quarkonium Working Group], 
  \href{https://link.springer.com/article/10.1140/epjc/s10052-010-1534-9}{Eur. Phys. J. C {\bf 71}, 1534 (2011)}.
  


  \bibitem{qcdme1} K. Gottfried, \href{https://journals.aps.org/prl/abstract/10.1103/PhysRevLett.40.598}{Phys. Rev. Lett. {\bf 40}, 598 (1978)}.
  
   \bibitem{qcdme2} M. B. Voloshin, \href{https://www.sciencedirect.com/science/article/abs/pii/0550321379900373?via\%3Dihub}{Nucl. Phys. B {\bf 154}, 365 (1979)}.

   \bibitem{qcdme3} T. M. Yan, \href{https://journals.aps.org/prd/abstract/10.1103/PhysRevD.22.1652}{Phys. Rev. D {\bf 22}, 1652 (1980)}.

   \bibitem{qcdme4} Y. P. Kuang and T. M. Yan, \href{https://journals.aps.org/prd/abstract/10.1103/PhysRevD.24.2874}{Phys. Rev. D {\bf 24}, 2874 (1981)}.

   \bibitem{qcdme5} Y. P. Kuang, \href{https://link.springer.com/article/10.1007/s11467-005-0012-6}{Front. Phys. China {\bf 1}, 19 (2006)}.

 \bibitem{PRD054019}
Q. Lu and Y. P. Kuang, 
 \href{https://journals.aps.org/prd/abstract/10.1103/PhysRevD.75.054019}
{Phys. Rev. D \textbf{75}, 054019 (2007).}

 \bibitem{PRD074033}
M. B. Voloshin, 
 \href{https://journals.aps.org/prd/abstract/10.1103/PhysRevD.86.074033}
{Phys. Rev. D \textbf{86}, 074033 (2012).}
      
 \bibitem{PRD012002}
M. Ablikim \textit{et al.} [BESIII Collaboration], 
 \href{https://journals.aps.org/prd/abstract/10.1103/PhysRevD.87.012002}
{Phys. Rev. D \textbf{87}, 012002 (2013).}

 \bibitem{PRD112018}
M. Ablikim \textit{et al.} [BESIII Collaboration], 
 \href{https://journals.aps.org/prd/abstract/10.1103/PhysRevD.91.112018}
{Phys. Rev. D \textbf{91}, 112018 (2015).}

\bibitem{CPC023001}
M. Ablikim \textit{et al.} [BESIII Collaboration], 
~\href{https://iopscience.iop.org/article/10.1088/1674-1137/42/2/023001}
{Chin. Phys. C \textbf{42}, 023001 (2018).}

 \bibitem{JHEP003}
M. Ablikim \textit{et al.} [BESIII Collaboration], 
 \href{https://link.springer.com/article/10.1007/JHEP05(2022)003}
{JHEP \textbf{05}, 003 (2022).}


\bibitem{psi2S2021}
M. Ablikim \textit{et al.} [BESIII Collaboration], 
~\href{https://iopscience.iop.org/article/10.1088/1674-1137/ad595b}
{Chin. Phys. C \textbf{48}, 093001 (2024).}

\bibitem{pdg2024} 
S. Navas \textit{et al.} [Particle Data Group], 
\href{https://journals.aps.org/prd/abstract/10.1103/PhysRevD.110.030001}
{Phys. Rev. D \textbf{110}, 030001 (2024).}


  \bibitem{Ablikim:2009aa} M. Ablikim {\it et al.} [BESIII Collaboration], \href{https://doi.org/10.1016/j.nima.2009.12.050}{Nucl. Instrum. Meth. A {\bf 614}, 345 (2010)}.

  \bibitem{CXYu_bes3} C.~H.~Yu {\it et al.}, Proceedings of IPAC2016, Busan, Korea (JACoW, Busan, 2016), \url{https://accelconf.web.cern.ch/ipac2016/}.


  \bibitem{Ablikim:2019hff} M.~Ablikim {\it et al.} [BESIII Collaboration], \href{https://iopscience.iop.org/article/10.1088/1674-1137/44/4/040001}{Chin. Phys. C {\bf 44}, 040001 (2020)}.

  \bibitem{EcmsMea}
  J.~Lu, Y.~Xiao, and X.~Ji, \href{https://doi.org/10.1007/s41605-020-00188-8}{Radiat. Detect. Technol. Methods {\bf 4}, 337(2020)}.

  \bibitem{EventFilter}
  J.~W.~Zhang {\it et al.}, \href{https://doi.org/10.1007/s41605-022-00331-7}{Radiat. Detect. Technol. Methods {\bf 6}, 289 (2022)}.


  \bibitem{tof_a} X. Li \textit{et al.}, \href{https://link.springer.com/article/10.1007\%2Fs41605-017-0014-2}{Radiat. Detect. Technol. Meth. {\bf 1} 13 (2017)}.


  \bibitem{tof_b} Y. X. Guo \textit{et al.}, \href{https://link.springer.com/article/10.1007\%2Fs41605-017-0012-4}{Radiat. Detect. Technol. Meth. {\bf 1}, 15 (2017)}.

  \bibitem{tof_c} P. Cao \textit{et al.}, \href{https://www.sciencedirect.com/science/article/pii/S0168900219314068?via\%3Dihub}{Nucl. Instrum. Meth. A {\bf 953}, 163053 (2020)}.

   \bibitem{geant4} S.~Agostinelli \textit{et al.} [GEANT4 Collaboration], \href{https://doi.org/10.1016/S0168-9002(03)01368-8}{Nucl. Instrum. Meth. A {\bf506}, 250 (2003).}


   \bibitem{kkmc_a} S. Jadach, B. F. L. Ward, and Z. Was, \href{https://www.sciencedirect.com/science/article/pii/S0010465500000485?via\%3Dihub}{Comput. Phys. Commun. {\bf 130}, 260 (2000)}.


    \bibitem{kkmc_b} S. Jadach, B. F. L. Ward, and Z. Was, \href{https://journals.aps.org/prd/abstract/10.1103/PhysRevD.63.113009}{Phys. Rev. D {\bf 63}, 113009 (2001)}.


    \bibitem{evtgen_a} R. G. Ping, \href{https://iopscience.iop.org/article/10.1088/1674-1137/32/8/001}{Chin. Phys. C {\bf 32}, 599 (2008)}.

   \bibitem{evtgen_b} D. J. Lange, \href{https://www.sciencedirect.com/science/article/pii/S0168900201000894?via\%3Dihub}{Nucl. Instrum. Meth. A {\bf 462}, 152 (2001)}.
 
 
\bibitem{pdg2022} 
R. L. Workman \textit{et al.} [Particle Data Group], 
\href{https://academic.oup.com/ptep/article/2022/8/083C01/6651666}
{PTEP \textbf{2022}, 083C01 (2022).}
   
    \bibitem{lundcharm_a} J. C. Chen, G. S. Huang, X. R. Qi, D. H. Zhang, and Y. S. Zhu, \href{https://journals.aps.org/prd/abstract/10.1103/PhysRevD.62.034003}{Phys. Rev. D {\bf 62}, 034003 (2000)}.

  \bibitem{lundcharm_b} R. L. Yang, R. G. Ping, and H. Chen, \href{https://iopscience.iop.org/article/10.1088/0256-307X/31/6/061301}{Chin. Phys. Lett. {\bf31}, 061301 (2014)}.


  \bibitem{photos} E. Barberio, B. van Eijk and Z. Was,  \href{https://www.sciencedirect.com/science/article/abs/pii/001046559190012A?via\%3Dihub} {Comput. Phys. Commun. {\bf 66}, 115 (1991). }

 \bibitem{PRD092009} 
M. Ablikim \textit{et al.} [BESIII Collaboration], 
\href{https://journals.aps.org/prd/abstract/10.1103/PhysRevD.86.092009}
{Phys. Rev. D \textbf{86}, 092009 (2012).}


 \bibitem{ref:fom} 
G. Punzi, 
\href{https://inspirehep.net/literature/634798}
{eConf C030908, MODT002 (2003).}

    
\bibitem{Johnson} N. L. Johnson, 
\href{https://doi.org/10.2307/2332539}
{Biometrika \textbf{36}, 149(1949).}

 \bibitem{PRD112005} 
M. Ablikim \textit{et al.} [BESIII Collaboration], 
\href{https://journals.aps.org/prd/abstract/10.1103/PhysRevD.83.112005}
{Phys. Rev. D \textbf{83}, 112005 (2011).}


 \bibitem{PRD052005}
M. Ablikim \textit{et al.} [BESIII Collaboration], 
 \href{https://journals.aps.org/prd/abstract/10.1103/PhysRevD.81.052005}
{Phys. Rev. D \textbf{81}, 052005 (2010).}

 \bibitem{PRL261801}
M. Ablikim \textit{et al.} [BESIII Collaboration], 
 \href{https://journals.aps.org/prl/abstract/10.1103/PhysRevLett.105.261801}
{Phys. Rev. Lett \textbf{105}, 261801 (2010).}

  \bibitem{YPG:bam} M. Ablikim $et$ $al$. [BESIII Collaboration],\textit{} \href{https://journals.aps.org/prd/abstract/10.1103/PhysRevD.87.012002}{Phys. Rev. D {\bf87}, 012002 (2013)}.


  \bibitem{barlow_test} O. Behnke, K. Kroninger, G. Schott, and T. H. Schorner-Sadenius, ``Data Analysis in High Energy Physics: A Practical Guide to Statistical Methods'', \href{https://www.wiley.com/en-cn/Data+Analysis+in+High+Energy+Physics\%3A+A+Practical+Guide+to+Statistical+Methods-p-9783527410583}{Wiley-VCH, Berlin, Germany, 2013}.

  
    \bibitem{arXiv:2408} M. Ablikim \textit{et al.} [BESIII Collaboration],  \href{http://arxiv.org/abs/2408.06677}{arXiv: hep-ex/2408.06677}.
    
  \bibitem{Stenson:2006gwf} K. Stenson, \href{https://arXiv.org/abs/physics/0605236}{arXiv: physics/0605236}.


  \bibitem{Liu:2015uha} X. X. Liu, X. R. Lyu, and Y. S. Zhu, \href{https://doi.org/10.1088/1674-1137/39/10/103001}{Chin. Phys. C {\bf 39}, 113001 (2015)}.



\end{thebibliography}
\end{document}